# AC magnetometry of van der Waals magnets using ultrasensitive Graphene Hall sensors

Eugene Park [a,b#], Jihoon Keum [a,b#], Ji-Hwan Baek [d], Hyuncheol Kim [a,b], Kenji Watanabe [e], Takashi Taniguchi [f], Gwan-Hyoung Lee [d], and Je-Geun Park [a,b,c]*

[a] Department of Physics and Astronomy, Seoul National University, Seoul 08826, South Korea

[b] Center for Quantum Materials, Department of Physics and Astronomy, Seoul National University, Seoul 08826, South Korea

[c] Institute of Applied Physics, Seoul National University, Seoul 08826, South Korea

[d] Department of Materials Science and Engineering, Seoul National University, Seoul 08826, Republic of Korea

[e] Research Center for Electronic and Optical Materials, National Institute for Materials Science, 1-1 Namiki, Tsukuba 305-0044, Japan

[f] Research Center for Materials Nanoarchitectonics, National Institute for Materials Science, 1-1 Namiki, Tsukuba 305-0044, Japan

# Equal contribution

*Corresponding Author: jgpark10@snu.ac.kr

# Abstract

Probing the dynamical magnetic properties of two-dimensional (2D) materials requires sensitive techniques capable of detecting small magnetic fields from nanoscale samples. We demonstrate quantitative AC and DC magnetometry of a ferromagnetic $Fe_{3-x}GeTe_2$ nanoflakes using ultrasensitive graphene Hall sensors. These devices achieve record-low magnetic field detection noise at both cryogenic and room temperature, enabled by hBN encapsulation, low-resistance fluorographene contacts, and a novel fabrication process. We perform quantitative AC susceptibility measurements up to 1 kHz, resolving both real and imaginary components with nanotesla-scale sensitivity and milliradian phase accuracy, the first such measurement in a van der Waals magnet. Our results establish graphene Hall sensors as a powerful and broadly applicable platform for studying magnetic and superconducting phases near the 2D limit.

# 1. Introduction

The advent of 2D van der Waals (vdW) magnets has brought renewed attention to nanoscale magnetism [1–3]. Quantifying physical quantities in such a small system is essential for understanding microscopic mechanisms and validating theoretical models. However, conventional magnetometry techniques often struggle to detect magnetic signals from nanoflakes due to their extremely small volume, typically on the order of 0.01 $\mu m^3$. This limitation has motivated the search for probes capable of operating in the nanoscale regime with both high sensitivity and broad applicability.

Magnetic moments, particularly their temperature, field, and frequency dependence, are crucial in characterizing magnetic and superconducting materials. Various probes have been developed (Figures 1a & S1), including magneto-optical Kerr effect (MOKE) [4–8], magneto circular dichroism (MCD) [7,9,10], Raman spectroscopy [2], scanning superconducting quantum interference device (SQUID) [11–13], and nitrogen-vacancy (NV) centers [14–16]. While each method has its strengths, they also suffer from important limitations: optical techniques are sensitive but not easily quantifiable, SQUIDs require cryogenic conditions [11,12], and NV centers have restricted field windows [14,15]. As a result, a wide parameter space at moderate temperatures and finite magnetic fields remains experimentally inaccessible (Figure 1a and Table S3).

Graphene Hall sensors offer a promising solution. Their high mobility [17], high Hall coefficient, low noise [18,19], and linear Hall response, particularly when encapsulated in hexagonal boron nitride (hBN) [20,21], allow sensitive magnetometry from cryogenic to room temperatures. Their 2D nature also makes them naturally compatible with vdW heterostructures. A previous study [22] has used such sensors for static field detection; however, their potential for probing dynamic magnetic responses remains unexplored, especially under realistic conditions involving finite temperatures, fields, and frequencies.

As the history of magnetism studies has demonstrated over the past half-century, the study of spin dynamics provides invaluable insights into the magnetic properties of the materials under discussion. For example, inelastic neutron scattering is a well-established tool for measuring spin dynamics in the high-frequency regime [23]. Further bulk probes, such as muon spin spectroscopy (µSR), nuclear magnetic resonance (NMR) and AC susceptibility (using conventional magnetic properties measurement systems) can probe sequentially lower frequency ranges (Figure S1) [24]. However, existing tools, which are well-suited for bulk systems, cannot be applied to 2D vdW magnets, even after considering ongoing technical developments, as the gauge volume of samples is far too small by several orders of magnitude.

Therefore, it is urgent to find an alternative solution for the future advancement of the field. If possible, this study will open a new window into capturing fluctuations, excitations, and relaxation processes in real time, yielding indispensable insight into the fundamental interactions that govern emergent quantum phenomena. The AC susceptibility is particularly informative and important. It directly measures information about spin relaxation, vortex dynamics, domain wall motion, and spin glass behavior, all of which are critical to understanding 2D correlated materials [24–33]. However, quantitative AC susceptibility measurement of vdW nanofalkes has not yet been achieved to date, particularly in terms of measuring both real ($\chi'$) and imaginary ($\chi''$) components with sufficient sensitivity and phase resolution.

In this work, we demonstrate quantitative AC and DC magnetometry of ferromagnetic $Fe_{3-x}GeTe_2$ (FGT) nanoflakes using ultrasensitive graphene Hall sensors. By combining ultraclean hBN encapsulation, low-resistance fluorographene contacts [34], and a new bubble-clearing transfer process, we achieve record magnetic field detection noise at both 4 and 300 K. Our device enables AC susceptibility measurements with nanotesla-scale sensitivity and milliradian phase accuracy up to 1 kHz, representing the first quantitative observation of $\chi''$ in a vdW magnet. These results establish graphene Hall sensors as a powerful platform for probing dynamic magnetic phenomena in 2D systems.

## 2. Results and Discussion

### 2.1 Fabricating Ultrahigh Sensitivity Graphene Sensors

We fabricated hBN-encapsulated graphene Hall sensors (Figures 1b-c), with a gold current line embedded on the side to generate an out-of-plane AC excitation magnetic field. The high-quality devices require ultraclean heterostructure interfaces with low contact resistance. To ensure clean interfaces and reproducible device performance, we developed a new fabrication protocol that enables the assembly of bubble-free hBN encapsulated graphene heterostructures (see Device Fabrication in the Supporting Information). This method employs a PVC/PDMS stamp [35–37] that has been modified to create microdomes, which act as soft springs to expel trapped bubbles during stacking at high temperatures [36]. This not only minimizes trapped bubbles at the interface but can also actively remove existing bubbles from the interface (Supplementary Video 1).

Furthermore, when compared to other polymers, the use of PVC offers superior adhesion [35, 37] at low temperatures and easy lift-off at high temperatures. The graphene channel and gold electrodes were patterned using e-beam lithography, followed by XeF$_2$ selective etching to form low-resistance 1D contacts [34]. This device structure enables accurate Hall resistance measurements across a wide range of temperatures and magnetic fields. All fabricated sensors showed a resistance peak at the charge neutrality point (CNP) (Figure 1d). Most devices exhibited high Hall coefficients (~5000 VA$^{-1}$T$^{-1}$, 300 K) and carrier mobility (~ 400,000 cm$^2$V$^{-1}$s$^{-1}$, 4 K). Additional data for other devices are provided in Figures S4-S8 and Table S2.

## 2.2 Noise Characteristics of Ultrahigh Sensitivity Graphene Sensors

The magnetic field detection noise is defined as $S_B^{1/2} = S_V^{1/2}/(IR_H)$, where $I$ is the current, $R_H$ is the Hall coefficient, and $S_V^{1/2}$ is the voltage power spectral density defined as $S_V(f) = \lim_{T \to \infty} \frac{2}{T}|F(f)|^2$, the time average of the Fourier amplitude squared. $S_B^{1/2}$ was evaluated across a range of back-gate voltages and drive currents. Achieving high magnetic-field sensitivity (i.e., low detection noise) requires both a large Hall coefficient and a low-noise device. Previous studies on solid-state Hall sensors have shown that 1/f noise dominates at low frequencies, especially in graphene devices [18–21,38–40]. Although the origin of the 1/f noise remains under debate, it is known to follow an area scaling of the normalized power spectral density $S_V^{1/2}/I \sim A^{-1/2}$ [39, 41], and is strongly affected by graphene–metal contact quality [42]. Our devices exhibited similar characteristics, with a dominant 1/f noise component at low frequency, as shown in Figure S9.

The $S_V^{1/2}$ depends on the source-drain current, doping level, and temperature. To quantify the magnetic field detection limits of our device, we measure the Hall coefficient and $S_V^{1/2}$ under various current and back-gate voltages at 4 and 300 K. Following the approach in Ref. 20, we extracted the magnetic field detection noise $S_V^{1/2}$ as a function of gate voltage and current (Supporting Information Section 2). The back-gate voltage dependence of the magnetic field detection noise for device D5 is shown in Figure 2a. Near the CNP at 0 V, the noise increases due to the low Hall coefficient and poor screening of charge fluctuations [19,20,42]. The best performance was achieved with device D5 near -1 V back-gate voltage, where we observed a minimum magnetic field detection noise at 4 K, being $S_B^{1/2} = 36 \pm 4 \text{ nTHz}^{-1/2}$ and at 300 K being $S_B^{1/2} = 76 \pm 8 \text{ nT Hz}^{-1/2}$, the lowest value reported for any solid-state Hall sensor to date.

We also evaluated the magnetic field detection noise under high-field conditions with the magnetic field in the out-of-plane direction (15 K, -9 T). We obtained $S_B^{1/2} = 1.9 \pm 0.2 \text{ μT Hz}^{-1/2}$, also the lowest value reported for solid-state Hall sensors in high magnetic fields. In this regime, we used $dR_{xy}/\mu_0 dH$ to estimate the magnetic field detection noise of the Hall coefficient (Figure S8). As shown in Figure 2c, the optimal gate voltage shifts to higher doping levels, likely due to the increased voltage noise, which has a large and broad peak near the CNP (Figure S10b). This increased voltage noise originates from the large and broad resistance near the CNP under large magnetic fields and high source-drain voltages (Figure S10a, Supporting Information), as this change dominates the gate voltage dependence over the Hall coefficient change near the CNP [43]. At high fields, charge fluctuations between localized and extended states in the quantum Hall regime also contribute to the overall decrease in sensitivity [20,44]. A more detailed discussion of the noise at high fields is provided in Section 3.1 of the Supporting Information. Although additional noise sources emerge at higher fields (Figure 2d), their magnitude remains similar to that of the 0 T case. This suggests that the primary factor contributing to the decreased sensitivity in magnetic field detection is the reduced Hall coefficient, as shown in Figure S10. These results suggest that further improvements in sensitivity under high-field conditions are possible by maximizing the Hall coefficient.

The performance of our devices is mainly limited by the Johnson noise at 1 kHz (Figure S11). In most cases, our ground noise level is relatively high, on the order of 30 nV Hz$^{-1/2}$ (Figure S9), which dominates the signal at higher frequencies. As a result, the magnetic field detection noise decreases with increasing current, implying that the measurement equipment has a dominant contribution to the noise. This behavior is illustrated in Figure 2b, where the detection noise decreases monotonically as the current increases. While the magnetic field detection limit could be further reduced at higher currents, we restrict the bias to 100 μA to avoid device damage. Current and voltage dependences for other devices are shown in Figures S4-S8. Assuming a lower ground noise level of 10 nV Hz$^{-1/2}$ at 4 K, $I = 100$ μA, $V_g = -1$ V of D5 (Figure S9) and extrapolating the $f^{-1/2}$ noise scaling to $f = 1$ kHz, the voltage noise $S_V^{1/2}$ could be reduced to $S_V^{1/2} \sim 10$ nV Hz$^{-1/2}$, corresponding to a magnetic field detection limit to $S_B^{1/2} \sim 12$ nT Hz$^{-1/2}$. Therefore, by reducing resistance in the measurement circuit, the magnetic field detection limit could be further pushed to only $\sim 10$ nT Hz$^{-1/2}$.

**2.3 DC and AC Micromagnetometry with Ultrahigh Sensitivity Graphene Sensors**

Using our high-sensitivity graphene sensors, we measured the magnetic properties of a vdW ferromagnet Fe$_{3-x}$GeTe$_2$ (FGT) [8, 45, 46]. A FGT nanoflake with a thickness of ~21 nm was placed on top of the hBN-encapsulated graphene heterostructure (Figure S3). During all measurements, the device was biased at a fixed gate voltage of $V_g = 4$ V, corresponding to an estimated carrier density of $n\sim0.26\times10^{12}$ cm$^{-2}$. For experiments under a DC magnetic field, an AC excitation current of 100 μA was applied to the graphene sensor, which elevated the electron temperature to suppress ballistic transport signatures [22] and simultaneously enhanced the Hall signal. Higher currents were avoided to minimize Joule heating.

Figure 3a shows hard ferromagnetic hysteresis loops in FGT, obtained after subtracting a linear Hall background from the raw data. The observed coercive fields are consistent with previous reports [47, 48]. As the temperature increases, both the coercive field and the loop width gradually decrease, disappearing completely above ~140 K. Figure 3b shows that near 130 K, the hysteresis loop adopts a qualitatively different shape marked by abrupt transitions from saturation to intermediate magnetization, with a nearly linear region in between. This behavior is attributed to the labyrinth domain structures in stoichiometric FGT [8].

The humps and small offsets in the positive and negative branches of the $R_{xy}$, the data in Figure 3a, which persist even above the magnetic transition temperature, are attributed to ballistic transport signatures [17,49,50]. These features arise when the cyclotron radius ($R_c = \frac{\hbar k_F}{B_c e}$) becomes comparable to the device channel width [49]. For device D1, the maximum inscribed radius of the channel is approximately 1.45 μm. Using $|k_F| = \sqrt{\pi n}$ in graphene, we estimate a critical field of $B_c \sim 0.2$ kOe, consistent with the hump location near ~0.2 kOe. When the coercive field is smaller than $B_c$, such ballistic features are absent, as shown in Figure 3b.

To quantitatively extract the magnetic moment of the FGT nanoflake, we establish a connection between its magnetization and the measured Hall voltage. Assuming the nanoflake is uniformly magnetized along the out-of-plane direction, the resulting magnetic field distribution was simulated using finite element modeling (Figure 3c). The magnetic field at the center of the graphene sensor is estimated to be ~ 10 G. Previous studies on ballistic Hall magnetometry[22,50,51] have demonstrated that the Hall voltage is directly proportional to the average magnetic field over the *flux-sensitive area* (FSA), the region between the transverse terminals (Figure S2a). This relationship can be expressed $V_{xy} = IR_H \langle B \rangle$, where $\langle B \rangle$ is the average magnetic field over the FSA. Numerically evaluating $\langle B \rangle$ allows us to define a conversion factor $\alpha$ (V/(A/m)) that links magnetization M to the Hall voltage. Using the hard ferromagnetic properties of FGT, we determine the saturated magnetization from the loop width via $\Delta V_{xy} = I(R_H(+M) - R_H(-M)) = IR_H\alpha(2M_{sat})$. Possible sources of absolute deviation from the true value due to the definition of the FSA in this method are discussed in Supporting Information Section 3.3.

Figure 3d shows the magnetization values extracted at various temperatures. Fitting the data to the critical scaling law $M = M_0\left(1 - \frac{T}{T_c}\right)^\beta$ yields $T_c = 120.0 \pm 0.1$ K, $\beta = 0.24 \pm 0.03$, and $M_0 = 1.06 \pm 0.03\ \mu_B\ \text{Fe}^{-1}$, which agrees well with the bulk saturation magnetization, $M_{sat} \sim 1.1\ \mu_B\ \text{Fe}^{-1}$ (Figure S13). The reduced transition temperature compared to the bulk value (~180 K) is likely due to the effects of exfoliation and the proximity to the near 2D limit [8,45,46]. The scaling exponent $\beta = 0.24 \pm 0.03$ is also consistent with a previous result for similarly thick samples, where $\beta \sim 0.25$ was reported [8]. In addition, we performed an independent DC measurement at a fixed magnetic field (Figure S14). Although this method does not allow for quantitative magnetization conversion, it reveals a bifurcation between the zero-field-cooled and field-cooled curves, a behavior consistent with prior reports on stoichiometric FGT [53].

We now turn to the measurement of AC susceptibility in FGT nanoflakes, using the scheme depicted in Figure 1b. An AC excitation magnetic field of ~0.225 G was applied during the measurement. Further details on the data processing procedures are available in Section 3.4 of the Supporting Information. We emphasize that, unlike superconductors, the induced magnetic field from a ferromagnet is several orders of magnitude smaller than the applied field, making this measurement a stringent benchmark of sensor sensitivity.

Figure 4a presents the real part ($\chi'$) of the AC susceptibility measured at different frequencies, showing a pronounced peak near 130 K. The peak value reaches ~2 $m\mu_B Fe^{-1} Oe^{-1}$, corresponding to an induced magnetic field of ~500 nT. A subtle but consistent frequency dependence is also observed across all datasets. On the other hand, the imaginary part ($\chi''$), shown in Figure 4b, has a peak value of ~0.5 $m\mu_B Fe^{-1} Oe^{-1}$ (~100 nT). A signal of this magnitude would be extremely challenging to detect with conventional magnetometry techniques. To overcome technical limitations, we developed a phase-fitting algorithm that models the lock-in phase sweep as a sinusoidal function. This approach significantly improved the phase accuracy by an order of magnitude in both noise simulations (Figure S15) and experimental data (Figure S16), as detailed in Section 3.2 of the Supporting Information. As a result, we achieved a phase sensitivity of ~1 mrad in the $\chi''$ measurement. This methodology can be universally applied to lock-in techniques. To the best of our knowledge, this constitutes the first quantitative measurement of the imaginary part of AC susceptibility in a magnetic vdW nanoflake. This unambiguously demonstrates that graphene Hall sensors reported here provide a powerful platform that makes spin dissipation directly accessible in magnetic van der Waals nanostructures. Measurements were performed up to 1 kHz: at higher frequencies, the device's intrinsic impedance introduces temperature-dependent phase shifts, limiting the dynamic range to below 1 kHz (Figure S17). Lower frequencies are not affected by this issue.

The observed peaks in both the real and imaginary components of the AC susceptibility are attributed to domain wall motion, a common feature in ferromagnets [8,32,54]. Consistent with this interpretation, our DC M-H measurements display anomalous hysteresis loops, which are linked to the formation of labyrinth-like domains near the transition temperature [8]. Then, domain wall motion in this phase will be responsible for the peak in the AC susceptibility. In this scenario, the imaginary part of the susceptibility arises from dissipation associated with domain wall dynamics [24]. Although the frequency dependence of the peak at the transition temperature is subtle, it may reflect slow spin dynamics, as previously reported in tellurium-rich FGT [30]. The low-temperature frequency dependence may be a combined effect of slow magnetic dynamics and the frequency-dependent measurement background. The possible slow spin dynamics and relatively large dissipation compared to other vdW magnets [26, 32] make (underdoped) FGT an exotic platform for further investigation of microscopic domain wall motion and glassy dynamics. A more detailed investigation of the frequency and field dependence of AC susceptibility is necessary to understand the microscopic mechanisms at full play. Nevertheless, our results demonstrate that graphene Hall sensors provide a powerful platform for probing magnetization dynamics in the sub-kHz frequency range. Such dynamical quantities furnish unique insights into spin dynamics, enabling access to microscopic degrees of freedom that remain inaccessible in static measurements.

We also performed AC susceptibility measurements while sweeping a DC field. In this configuration, the measured signal is proportional to the differential magnetization $dM/\mu_0 dH$, corresponding to the slope of the M-H curve [24]. For a hard ferromagnet such as FGT, the differential magnetization becomes peaked, approaching a delta function, at the magnetization reversal, which manifests as a sharp peak in the AC susceptibility (Figure 4d). Once the magnetization switches, it becomes irreversible at the same DC field, resulting in an instantaneous change and rendering the AC phase ill-defined. The broad hump centered near 0 kOe is attributed to an increase in the Hall coefficient from ballistic effects, consistent with the hump ending near $B_c \sim 0.2$ kOe, as seen in Figure 3a. The linear background likely originates from the differential magnetoresistance in the graphene layer, due to the longitudinal offset between the two transverse terminals. The position of the coercive peaks shifts with temperature, decreasing at higher temperatures and disappearing above the transition temperature. In Figure 4d, we compare coercive fields extracted from both DC and AC measurements, confirming that AC susceptibility can readily probe magnetic switching even under finite DC bias fields.

## 2.4 Feasibility of Ultrahigh Sensitivity Graphene Sensors

To demonstrate the robust feasibility of ultrahigh sensitivity required for AC measurements of magnetic nanoflakes, we fabricated and characterized 6 devices, and benchmarked their performance against previous reports [20,21,39,40,55–67]. An FGT nanoflake was added to one of these devices to perform the AC magnetometry presented. This comparison, summarized in Table S1, accounts for the scaling $S_V^{1/2}/I \sim A^{-1/2}$ and is presented in Figure 5. Even after normalizing for channel area using $(AS_B)^{1/2}$, our devices surpass the previous room-temperature record (Figure 2e, labeled hBN) [21] by a factor of 2.5, and are comparable to the best-performing graphite dual-gated devices at 4 K [20] without the need for proximal graphite gates. Additionally, our magnetic field detection limit at high fields greatly outperforms previous benchmarks [20]. The high sensitivity attainable with only hBN encapsulation represents a significant merit for scalable fabrication, while still allowing further enhancement through optimized device geometry. We attribute this improved sensitivity to the use of XeF$_2$ etched fluorographene contacts, which provide robust and ultralow contact resistance, as well as to the ultraclean encapsulation afforded by hBN layers, both of which offer reproducible and scalable fabrication advantages.

## 3. Conclusion

Our results establish ultrasensitive hBN-encapsulated graphene Hall sensors as a uniquely sensitive and broadly applicable platform for probing both static and dynamic magnetic properties of nanoscale vdW magnets. With the graphene Hall sensors made through the robust fabrication process using a new bubble cleaning method and XeF$_2$ etched contacts, we demonstrated a record-breaking sensitivity at 0 T of magnetic field, $S_B^{1/2} = 36 \pm 4 \text{ nT Hz}^{-1/2}$ at 4 K and $S_B^{1/2} = 76 \pm 8 \text{ nT Hz}^{-1/2}$ at 300 K. We further demonstrated its highly sensitive magnetic field detection limit of $S_B^{1/2} = 1.9 \pm 0.2 \text{ μT Hz}^{-1/2}$ at high fields (-9 T). Using these highly sensitive Hall sensors, we were able to measure the DC magnetization and the AC susceptibility of an FGT nanoflake.

Compared to existing magnetometry techniques, our devices offer a rare combination of quantitative sensitivity, broad operating conditions, and compatibility with van der Waals systems. For instance, NV center magnetometry offers excellent spatial resolution and magnetic sensitivity, but its operation is limited to low magnetic fields and requires specialized cryogenic optics. Scanning SQUIDs achieve exceptional field resolution but require cryogenic temperatures and complex instrumentation. On the other hand, optical techniques such as MOKE and MCD are convenient but generally lack quantitative accuracy, especially for measuring volume magnetization. In contrast, our graphene Hall sensors operate across a wide range of temperatures, from cryogenic to room temperatures, and over a broad magnetic field range, making them highly versatile. Furthermore, their 2D geometry and small footprint make them naturally suited to integration with exfoliated flakes and heterostructures.

Moreover, our approach offers robust compatibility with variable backgrounds, including finite magnetic fields and frequency-dependent excitation. We demonstrated this by measuring differential susceptibility under sweeping DC fields, obtaining consistent results with DC magnetization measurements. This unique field-tolerant feature of our device fills a key gap that has been left open by existing nanoscale magnetometry methods. It makes our platform suitable for studying systems with hysteresis, metamagnetic transitions, or field-tuned phase boundaries.

Looking forward, graphene Hall micromagnetometry offers broad opportunities for investigating slow dynamics and dissipation in two-dimensional quantum materials. These include frustrated magnets, skyrmion-hosting systems, vortex dynamics in 2D superconductors, and proximity-induced phenomena in hybrid heterostructures. With further improvements in sensor geometry and signal conditioning, the technique can be extended to higher frequencies, smaller samples, or time-resolved measurements. We anticipate that graphene Hall sensors will become a powerful and general-purpose tool in the evolving toolkit for mesoscopic magnetism and quantum materials research.

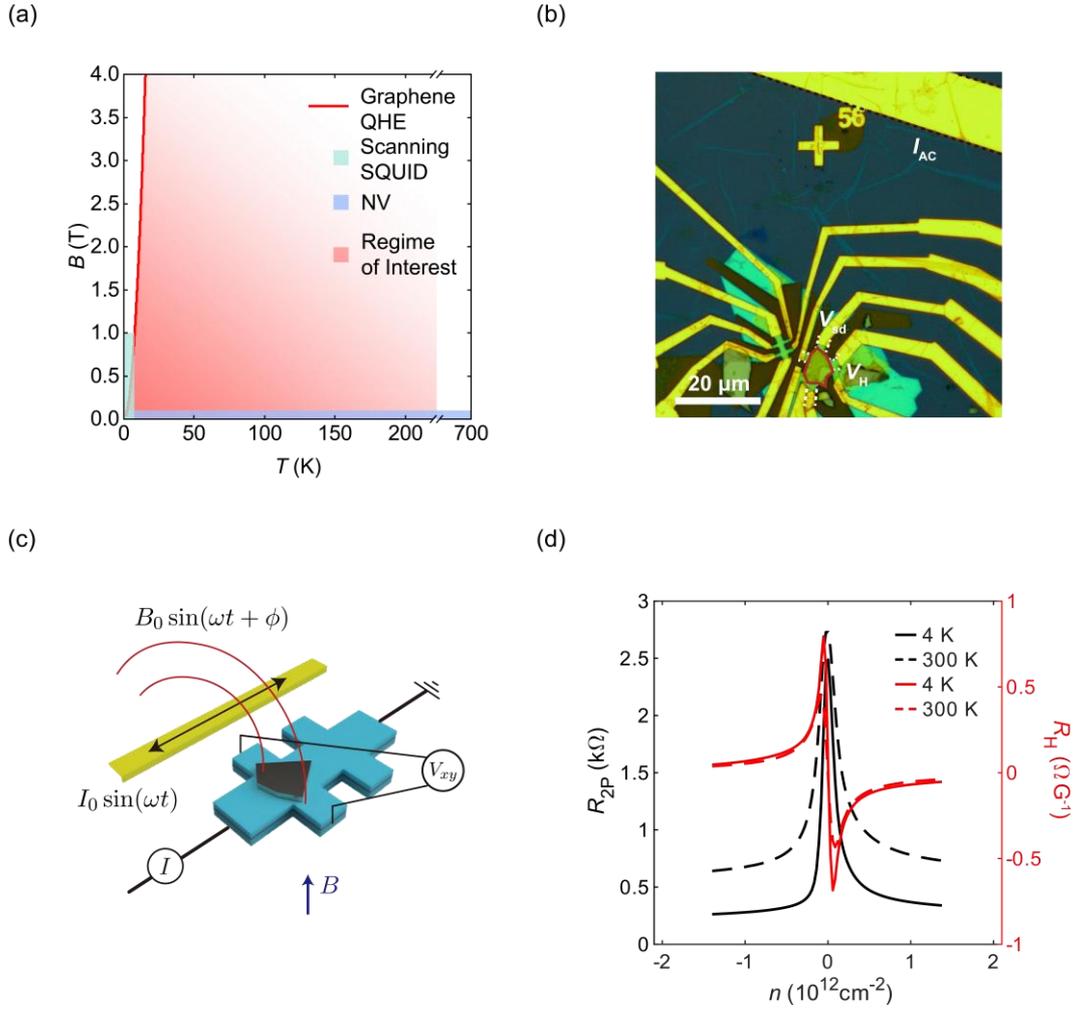

**Figure 1.** (a) Comparison of probes capable of quantifying magnetic fields based on magnetic field ($B$) and temperature ($T$), with the target of our device depicted as a wide red area in the B-T phase space. The bounds for scanning SQUID (teal) and NV (sky blue) are organized in Table S3 and Figure S1. The quantum Hall effect plateaus in the transverse voltage put an ultimate barrier for measurements using Hall sensors. (b) Optical microscope image of the graphene device. The boundary of the $Fe_{3-x}GeTe_2$ sample is marked in red, and the white dashes are guides for the source, drain, and transverse terminals. (c) Device geometry with measurement schematic. The magnetic field ($B$) is applied in the out-of-plane direction, defined as positive when pointing in the direction from the substrate to the sensor. (d) 2-point resistance and Hall coefficient of graphene device D1 at 4 and 300 K with an AC source of 100 μA used for the measurement. The electron carrier density is denoted as $n$. For other devices, see Figures S4-S8.

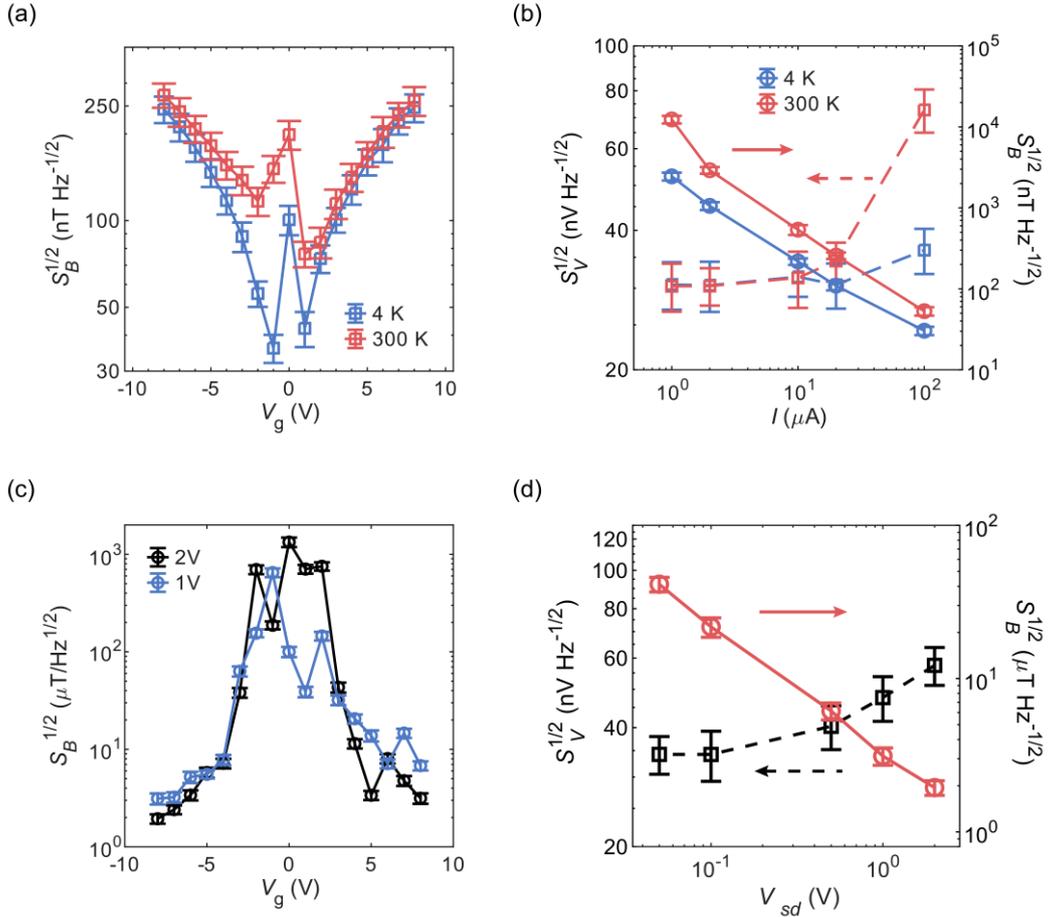

**Figure 2.** (a) The magnetic field detection noise ($S_B^{1/2}$) dependence by gate voltage ($V_g$) at an external magnetic field of 0 T and temperatures 4 and 300 K for D5. A constant current of 100 µA was used, which yielded the lowest magnetic field detection noise. (b) Current dependence of the voltage noise ($S_V^{1/2}$) and the magnetic field detection noise ($S_B^{1/2}$) of D5 at an external magnetic field of 0 T for temperatures 4 and 300 K. The gate voltage ($V_g$) was kept to yield minimum magnetic field noise in (a) at each temperature, $V_g = -1$ V for 4 K and $V_g = 1$ V for 300 K. (c) The magnetic field detection noise ($S_B^{1/2}$) dependence on gate voltage ($V_g$) at an external magnetic field of -9 T and a temperature of 15 K for D10. The applied voltages depicted correspond to a constant current of 2 V: 40 µA and 1 V: 20 µA. (d) Current dependence of the voltage noise ($S_V^{1/2}$) and the magnetic field detection noise ($S_B^{1/2}$) of D10 at an external magnetic field of -9 T and a temperature of 15 K. The gate voltage ($V_g = -7$ V) was kept to yield minimum magnetic field noise in (c).

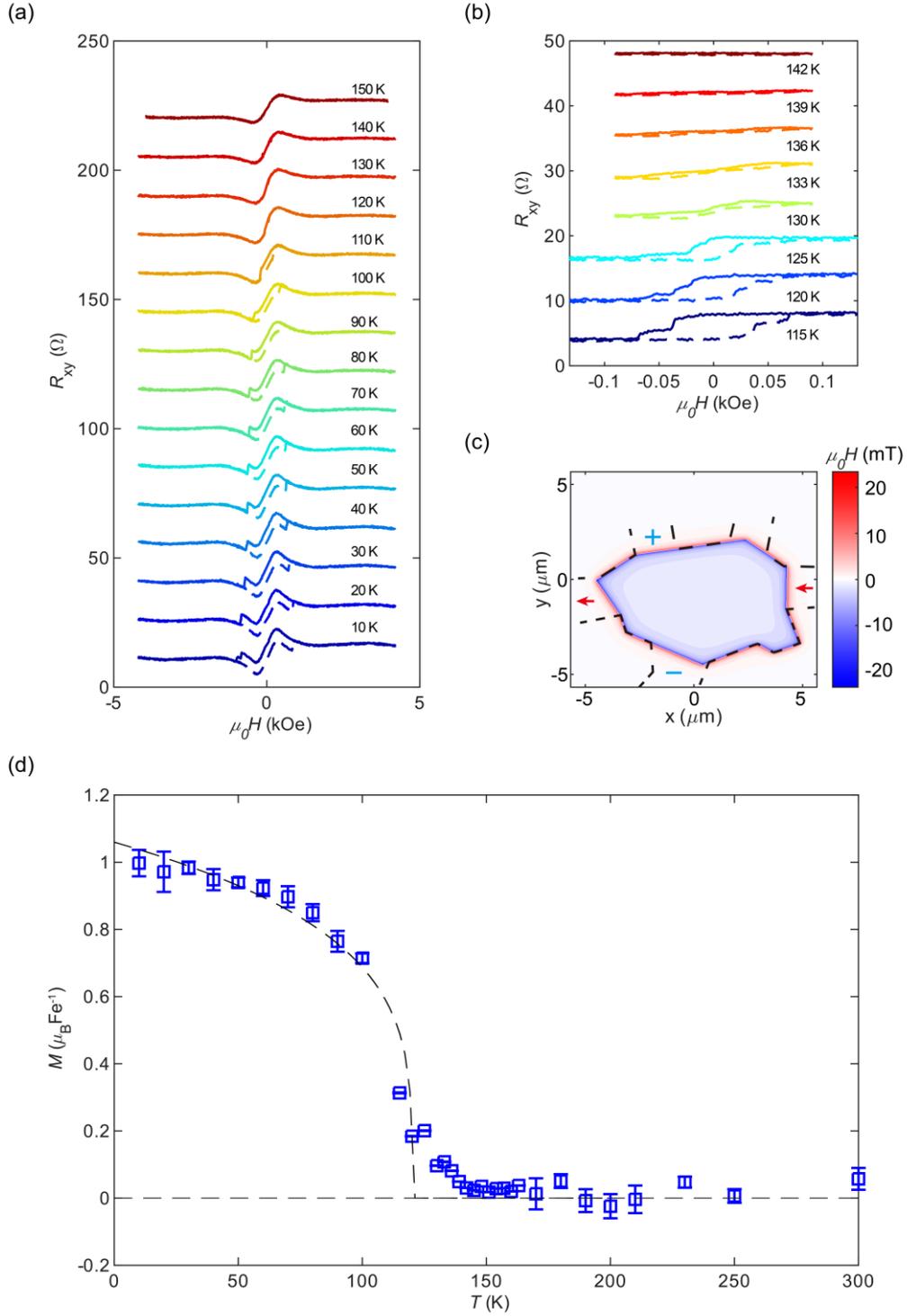

**Figure 3.** (a) The Hall resistance for D1 with $Fe_{3-x}GeTe_2$ at different temperatures. The hump appears at approximately 0.2 kOe. An AC current of 100 μA was used for the measurement. (b) The Hall resistance data for D1 with $Fe_{3-x}GeTe_2$ near the transition temperature $T_c$. (c) Finite element simulation of the out-of-plane magnetic field at the graphene plane due to the $Fe_{3-x}GeTe_2$ nanoflake. The magnetic field at the center of the flake is approximately ~10 G. The graphene boundary and the source-drain current (red arrows) with the transverse leads (+,

-) are illustrated. (d) The magnetization of the Fe$_{3-x}$GeTe$_2$ nanoflake shown as a function of temperature. The magnetization values are calculated from the hysteresis loops. The fit to $M = M_0 \left(1 - \frac{T}{T_c}\right)^\beta$ yields $T_c = 120.0 \pm 0.1$ K , $\beta = 0.24 \pm 0.03$ , and $M_0 = 1.06 \pm 0.03$ $\mu_B$ Fe$^{-1}$.

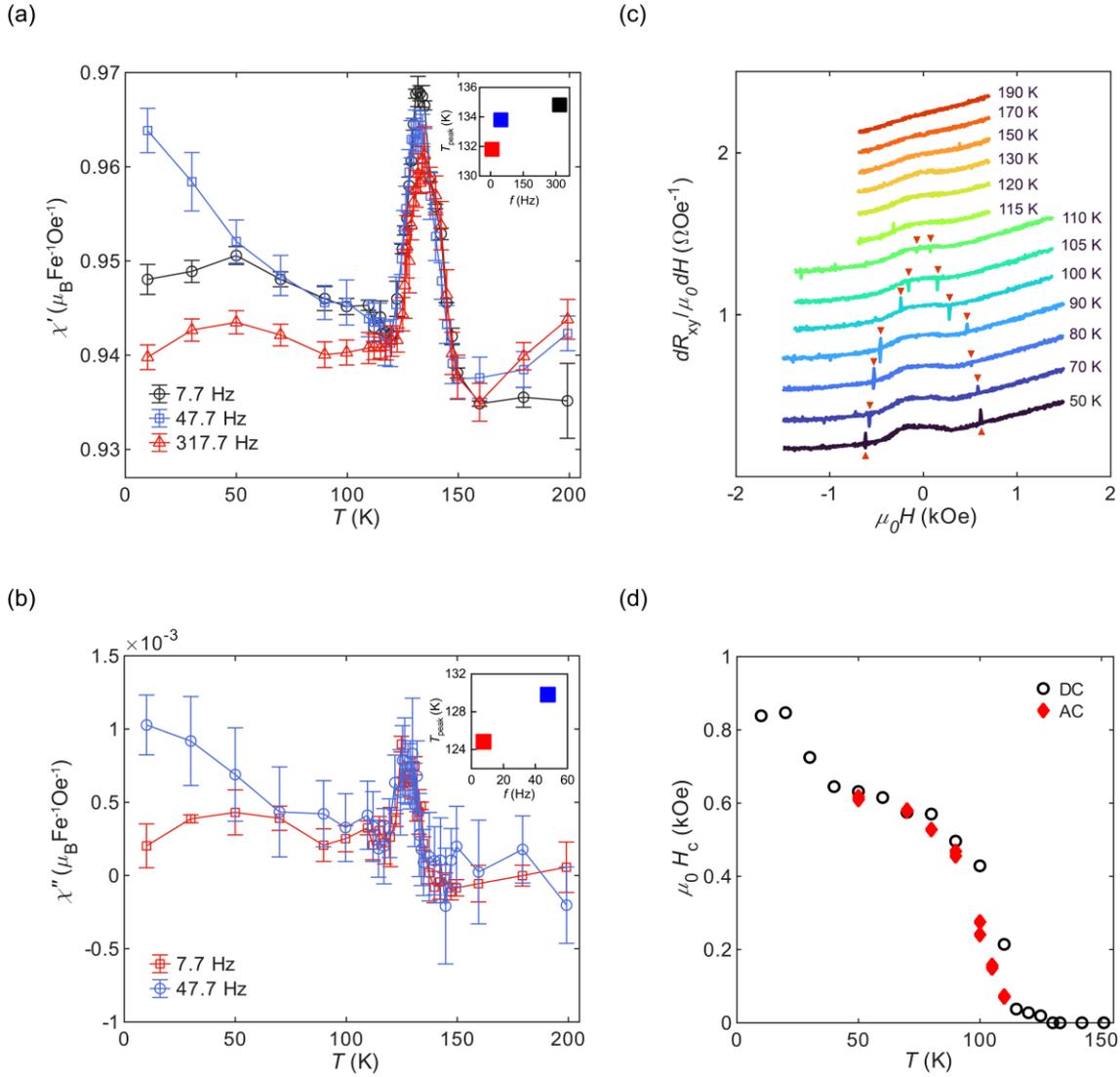

**Figure 4.** (a) The real part ($\chi'$) of the AC susceptibility of the $Fe_{3-x}GeTe_2$ nanoflake at different temperatures with an inset showing the peak position ($T_{peak}$) by frequency for different excitation field frequencies. Frequencies reaching 1 kHz are shown in Figure S17. The peak appears at $\approx 130$ K. The AC excitation field gives a constant offset to the data. (b) The imaginary part ($\chi''$) of the AC susceptibility for $Fe_{3-x}GeTe_2$ flakes with an inset showing the peak position ($T_{peak}$) by frequency by frequency. The peak position coincides with the real part of the signal. (c) The differential Hall resistance is measured while sweeping the background DC field. The red arrows indicate abrupt peaks resulting from magnetization reversals. (d) Comparison of the AC susceptibility magnetization reversals and the magnetization reversals in the M-H loops of Figure 3a. The AC excitation magnetic field for all measurements was 0.225 G.

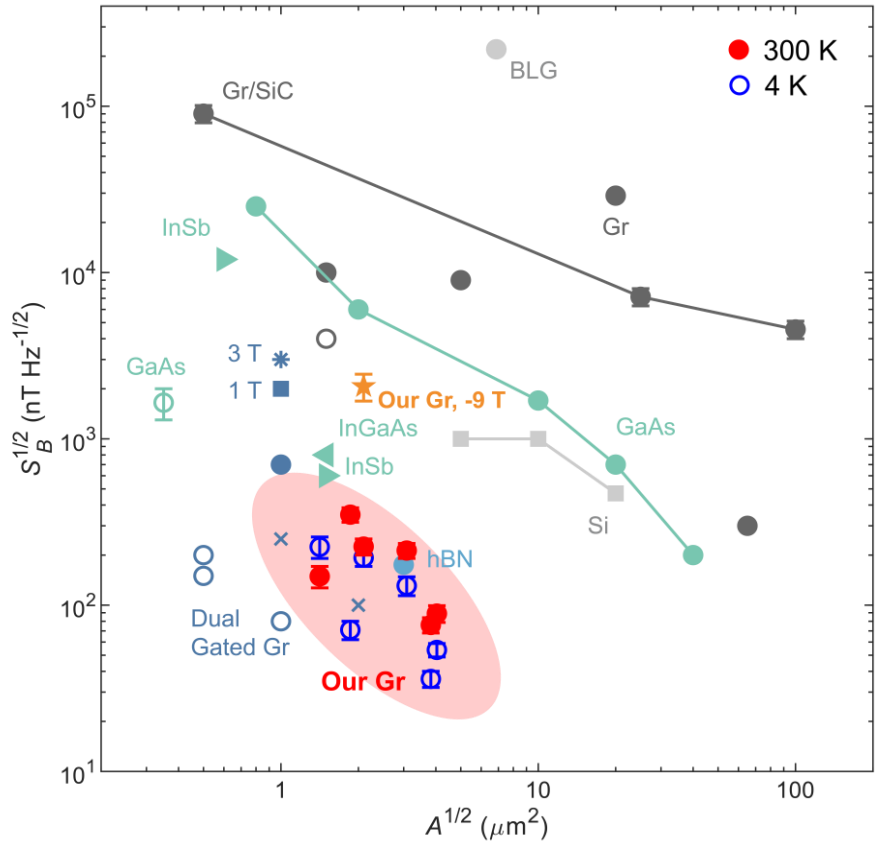

**Figure 5.** Comparison of the magnetic field detection noise of our devices with previously reported Hall sensors based on channel area ($A$). Data obtained from different schemes are marked with different color and symbol types. The open (filled) circles denote a measurement temperature of 4 K (300 K). Labels indicate the sensor material, with the following abbreviations: Bilayer graphene (BLG), Graphene (Gr), hBN-encapsulated graphene (hBN). Dual-gated graphene devices (graphite gates: circles, metal gates at 4 K: x) and graphite-gated devices at 1 T, 3 T are the same navy color. Our device, when measured at high fields, is the orange star. The extracted values from the preceding reports used for the comparison are organized in Table S1.

## Data availability statement

All data that support the findings of this study are included within the article (and any supplementary files).

## Acknowledgements

The work at CQM and SNU was supported by the Samsung Science & Technology Foundation (Grant No. SSTF-BA2101-05). One of the authors (J.-G.P.) was partly funded by the Leading Researcher Program of the National Research Foundation of Korea (Grant No. 2020R1A3B2079375). G.H.L. was financially supported through the core center program (2021R1A6C101B418) by the Ministry of Education and acknowledges the support from the Research Institute of Advanced Materials (RIAM), Institute of Engineering Research (IER), Institute of Applied Physics (IAP), SOFT Foundry Institute, and the Inter-University Semiconductor Research Center (ISRC) at Seoul National University. K.W. and T.T. acknowledge support from the JSPS KAKENHI (Grant Numbers 21H05233 and 23H02052) , the CREST (JPMJCR24A5), JST and World Premier International Research Center Initiative (WPI), MEXT, Japan.